\documentclass{article}
\usepackage{spconf,amsmath}

\usepackage{mediabb}
\usepackage{color}
\usepackage{multirow}
\usepackage{algorithm,algorithmic}
\usepackage{url}
\usepackage{amssymb,amsmath,bm}
\usepackage{textcomp}
\usepackage{cite}
\usepackage{CJKutf8}
\usepackage{hhline}
\usepackage{cite}
\usepackage{bbding}
\usepackage{pifont}
\usepackage{wasysym}

\def\L{{\cal L}}

\newcommand{\argmax}{\mathop{\rm arg~max}\limits}
\newcommand{\maru}[1]{\raise0.2ex\hbox{\textcircled{\scriptsize{#1}}}}

\title{Improving Scheduled Sampling for Neural Transducer-based ASR}
\name{\begin{tabular}{c} Takafumi Moriya, Takanori Ashihara, Hiroshi Sato, Kohei Matsuura, Tomohiro Tanaka, Ryo Masumura \end{tabular}}
\address{\begin{tabular}{c} NTT Corporation, Japan \end{tabular}}

\begin{document}
\ninept
\maketitle
\begin{abstract}
  The recurrent neural network-transducer (RNNT) is a promising approach for automatic speech recognition (ASR)
  with the introduction of a prediction network that autoregressively considers linguistic aspects. 
  To train the autoregressive part, the ground-truth tokens are used as substitutions for the previous output token,
  which leads to insufficient robustness to incorrect past tokens; a recognition error in the decoding leads to further errors.
  Scheduled sampling (SS) is a technique to train autoregressive model robustly to past errors
  by randomly replacing some ground-truth tokens with actual outputs generated from a model.
  SS mitigates the gaps between training and decoding steps, known as exposure bias,
  and it is often used for attentional encoder-decoder training. 
  However SS has not been fully examined for RNNT because of the difficulty in applying SS to RNNT due to the complicated RNNT output form. 
  In this paper we propose SS approaches suited for RNNT. 
  Our SS approaches sample the tokens generated from the distiribution of RNNT itself, i.e. internal language model or RNNT outputs.
  Experiments in three datasets confirm 
  that RNNT trained with our SS approach achieves the best ASR performance.
  In particular, on a Japanese ASR task, our best system outperforms the previous state-of-the-art alternative.
\end{abstract}
%

%\vspace{-0.05cm}
\section{Introduction}
\label{sec:intro}
%\vspace{-0.15cm}
End-to-end automatic speech recognition (E2E-ASR), which can directly map acoustic features to output tokens,
is attracting attention in the ASR research community. 
Several E2E-ASR modeling approaches have been proposed,
e.g. connectionist temporal classification (CTC)~\cite{Graves2006},
attentional encoder-decoder (AED)~\cite{chorowski2014,vaswani2017att} and recurrent neural network-transducer (RNNT)~\cite{Graves2012}.
In particular, RNNT is a promising technology, 
and has been reported in many recent
studies~\cite{erik2019drf,saon2020alsd,anmol2020conformer,Ehsan2020,kurata2020kdrnnt,zhou2021rnntcess,Cui2021rnntss,meng2021ilme,meng2021ilmt,zeyer21libritrans,karita2021csj,moriya2021simple,moriya2021rnnts2s,moriya2022rnntadlm,moriya2022tsrnnt,kubo2022kdlm2rnnt,zhou2022phonemernnt}.
In this paper, we focus on improving RNNT.

RNNT is composed of encoder, prediction, and joint networks.
The encoder and prediction networks process acoustic and linguistic information respectively,
and the resulting encoded features are fed to the joint network to obtain the prediction.
The unique characteristic of RNNT is that the introduction of the prediction network allows it to drop the assumption conditional independence
between predictions at different time steps.
This means that the RNNT outputs are jointly conditioned on not only acoustic information but also past linguistic information. 
Therefore, techniques for adding noise to the input can be applied to both encoder and prediction networks.
The acoustic masking approach known as SpecAugment~\cite{specaugment}, which randomly masks some acoustic features,
prevents overfitting the encoder network and improves ASR performance regardless of the E2E-ASR, i.e. CTC, AED, and RNNT. 
As a way to modify linguistic information, scheduling sampling (SS)~\cite{bengio2015ss} is a promising way to improve robustness for autoregressive model.
SS was originally proposed for AED, but it has yet to be fully examined for RNNT.
Thus in this paper, we focus on effectively applying SS for RNNT. 

RNNT and AED use ground truth tokens for the input to the decoder during training, 
but are prone to errors in inference step, i.e. exposure bias. 
SS randomly replaces some ground truth tokens with actual outputs of the training model. 
Thus tokens containing some errors are fed back to the training model, 
and the loss is computed using original ground truth tokens. 
This mitigates mismatch during inference. 
SS is often used in AED training but seldom in RNNT training.
This is because RNNT output shape is different from AED output due to the introduction of the prediction network, 
which complicates the application of SS approaches to RNNT training.

SS approaches for RNNT have been explored. %~\cite{zhou2021rnntcess,Cui2021rnntss}. 
In~\cite{zhou2021rnntcess}, they use frame-wise labels derived from external alignments generated by a hidden Markov model (HMM) hybrid system. 
The length of the labels equals that of acoustic features resulting in appropriate RNNT output length. 
Thus SS can, in the same way as AED training, be applied to RNNT training, 
but this requires additional efforts to obtain the alignments. 
In~\cite{Cui2021rnntss}, an external language model (ELM), which is pre-trained with only text data,
is utilized for sampling tokens and its length can match that of the ground truth tokens. 
This realizes the SS approach for RNNT training without external alignments,
and improved ASR performance. 
However the sampling is based on ELM output distribution, not on the RNNT itself. 
Their approach is similar to distilling the knowledge from ELM~\cite{kubo2022kdlm2rnnt}. 
Since we assume that the exposure bias problem remains unsolved, 
sampling should be performed using the output distribution of RNNT during training. 

In this paper, we propose two modifications to the SS approach in~\cite{Cui2021rnntss}. 
The first is to replace ELM with internal LM (ILM) which consists of prediction and joint networks of RNNT. 
The ILM parameters are fully shared with RNNT,
and thus the ILM outputs partly follow the RNNT output distribution. 
The other is to use the output distributions of the whole RNNT.
To make the length of RNNT output equal that of the ground truth tokens, 
we propose a sampling method which extracts the same number tokens as ground truth tokens from RNNT output. 
The time indices corresponding to each ground truth token for the extraction are determined by RNNT alignments during training, 
and thus the extracted tokens are conditioned on the distribution of whole RNNT. 
Therefore, we can apply the whole RNNT output to SS for RNNT training. 
We expect that both proposed SS variants will, using actual RNNT outputs, mitigate exposure bias.
Moreover, the SS approaches of ~\cite{bengio2015ss,Cui2021rnntss} take a token-by-token technique,
and so cannot process all utterances at once resulting in slow training. 
We also propose a proficiency-based utterance-level SS that samples each utterance.

We evaluate our proposed SS approaches on three datasets. 
The results demonstrate that our proposals consistently yield better ASR performance than RNNT using ELM with and without SS
even though our proposed SS variants do not need any external alignment or a pretrained ELM. 
In particular, on a Japanese ASR task, our best system outperforms the previous state-of-the-art alternative.

% Section 2
%\vspace{-0.35cm}
%\vspace{-0.1cm}
% Section 2
\section{ASR and Language Models}
\label{sec:modeling}
%\vspace{-0.15cm}
Let us first explain RNNT and LMs, which are the foundations of this work. 
Let $\bm{X} = \left[ \bm{x}_1, ..., \bm{x}_{T^{\prime}} \right]$ be the acoustic feature sequence, 
and $Y = \left[ y_1, ..., y_U \right]$ be a token sequence and $y_u \in \{1, ..., K\}$. 
$K$ is the number of tokens including special symbols, i.e. ``blank'' label, $\phi$, for RNNT,
and ``start/end-of-sentence'' labels for LMs.

%\vspace{-0.35cm}
\subsection{Recurrent neural network-transducer (RNNT)}
\label{sec:rnnt}
%\vspace{-0.2cm}
RNNT learns the mapping between sequences of different lengths. 
By introducing the prediction network, 
RNNT can consider the posterior probabilities to be jointly conditioned on not only encoder outputs but also previous predictions 
as shown in the following steps. 
First, $\bm{X}$ is downsampled and encoded into $\bm{H}^{\text{enc}} = \left[ \bm{h}^{\text{enc}}_{1}, ..., \bm{h}^{\text{enc}}_{T} \right]$
with length-$T$ via encoder network $f^{\text{enc}}(\cdot)$. 
Next, the tokens, $Y$, are also encoded into 
$\bm{H}^{\text{pred}} = \left[ \bm{h}^{\text{pred}}_{1}, ..., \bm{h}^{\text{pred}}_{U} \right]$ 
via prediction network $f^{\text{pred}}(\cdot)$. 
These encoded features are then fed to feed-forward network $f^{\text{joint}}(\cdot)$.
The above operations, which yield prediction $\hat{\bm{y}}_{t,u}$, are defined as follows:
%\vspace{-0.4cm}
\begin{eqnarray}
\bm{h}^{\text{enc}}_{t} &=& f^{\text{enc}} (\bm{x}_{t^{\prime}}; \theta^{\text{enc}}), \\
\bm{h}^{\text{pred}}_{u} &=& f^{\text{pred}} (y_{u-1}; \theta^{\text{pred}}), \\
\hat{\bm{y}}_{t,u} &=& \text{Softmax} \left(f^{\text{joint}} (\bm{h}^{\text{enc}}_{t}, \bm{h}^{\text{pred}}_{u}; \theta^{\text{joint}}) \ / \ Z \right),
%\end{aligned}\right.
\label{eq:rnnt}
\end{eqnarray}
where $\text{Softmax}(\cdot)$ means a softmax layer without learnable parameters,
and $Z$ is temperature used in inferencing. 
RNNT outputs three dimensional tensor $\hat{\bm{Y}}^{\text{RNNT}}_{1:T,1:U} \in \mathbb{R}^{T \times U \times K}$ from the training step.
The learnable parameters $\theta^{\text{RNNT}} \triangleq [\theta^{\text{enc}}, \theta^{\text{pred}}, \theta^{\text{joint}}]$
are optimized by using RNNT loss $\L_{\text{RNNT}}$ using the forward-backward algorithm~\cite{Graves2012}.

%\vspace{-0.3cm}
\subsection{Language models (LMs)}
%\vspace{-0.2cm}
In this work, we use two LMs, i.e. external language model (ELM) and internal language model (ILM),
for SS in the training step. We explain these below.

\subsubsection{External language model (ELM)}
ELM is a completely separate model from RNNT,
and trained with only transcriptions. 
ELM consists of language encoder ``lang-enc'' and linear output layer ``out'',
and outputs prediction $\hat{\bm{y}}_{u}$. 
The procedure is defined as follows:
\begin{eqnarray}
  \hat{\bm{y}}_{u}^{\text{ELM}} = \text{Softmax} \left( f^{\text{out}} (f^{\text{lang-enc}} (y_{u-1}; \theta^{\text{lang-enc}}); \theta^{\text{out}}) \right), 
\label{eq:elm}
\end{eqnarray}
where each parameter $\theta^{*}$ is learnable.
ELM outputs two dimensional matrix $\hat{\bm{Y}}^{\text{ELM}}_{1:U} \in \mathbb{R}^{U \times K}$ in the training step. 
Thus all parameters $\theta^{\text{ELM}} \triangleq [\theta^{\text{lang-enc}}, \theta^{\text{out}}]$ are optimized by using CE loss.

\subsubsection{Internal language model (ILM)}
ELM and ILM perform the same operation.
The difference is that ILM is a part of RNNT. 
ILM is defined as follows:
\begin{eqnarray}
  \hat{\bm{y}}_{u}^{\text{ILM}} = \text{Softmax} \left( f^{\text{joint}} (f^{\text{pred}} (y_{u-1}; \theta^{\text{pred}}); \theta^{\text{joint}}) \right),
\label{eq:ilm}
\end{eqnarray}
where all learnable parameters $\theta^{\text{ILM}} \triangleq [\theta^{\text{pred}}, \theta^{\text{joint}}]$ are fully contained in $\theta^{\text{RNNT}}$. 
The decoder inputs with acoustic information, i.e. $\bm{h}^{\text{enc}}_{t}$, are zeroed out.
ILM also outputs $\hat{\bm{Y}}^{\text{ILM}}_{1:U} \in \mathbb{R}^{U \times K}$, and its parameters are optimized by using CE loss $\L_{\text{ILM}}$ as in~\cite{Ehsan2020,meng2021ilmt}. 

\subsection{Training and decoding}
\vspace{-0.1cm}
\label{trainingdecoding}
\textbf{Training:}
This work follows our previous work~\cite{moriya2022rnntadlm}. 
We combine RNNT loss with CTC~\cite{Graves2006} and ILM~\cite{Ehsan2020,meng2021ilmt} losses as follows:
\vspace{-0.2cm}
\begin{equation}
\L = \L_{\text{RNNT}} + \alpha \L_{\text{CTC}} + \beta \L_{\text{ILM}},
\label{eq:trainloss}
%\vspace{-0.2cm}
\end{equation}
where $\alpha$ and $\beta$ are the weights of CTC and ILM losses, respectively.
$\L_{\text{CTC}}$ is CTC loss calculated between ground truth token sequence
and extra linear layer outputs $f^{\text{linear}}(f^{\text{enc}} (\bm{X}; \theta^{\text{enc}}); \theta^{\text{linear}})$.
Note that CTC extra linear layer $\theta^{\text{linear}}$ is thrown away during decoding.

\noindent \textbf{Decoding:}
We perform ILM estimation~\cite{meng2021ilme} if we use LM modules. 
This is explained as follows:
\vspace{-0.2cm}
\begin{eqnarray}
  \hat{Y} &=& \argmax_{Y} ( \text{log} \ p_{\text{RNNT}}(Y|\bm{X}) +  \mu_{1} \ \text{log} \ p_{\text{LM}}(Y) \nonumber \\
  &-& \mu_{2} \ \text{log} \ p_{\text{ILM}}(Y) + \mu_{3} \ |Y| ),
  \label{eq:hypilm}
%  \vspace{-0.2cm}
\end{eqnarray}
where $\mu_{*}$ are the weights of LM, ILM and length penalty $|Y|$ scores, respectively.
These are tuned using each development set. 

\begin{figure*}[t]
  \begin{center}
    %\vspace{-0.3cm}
    \hspace{0.2cm} \includegraphics[clip, width=17cm]{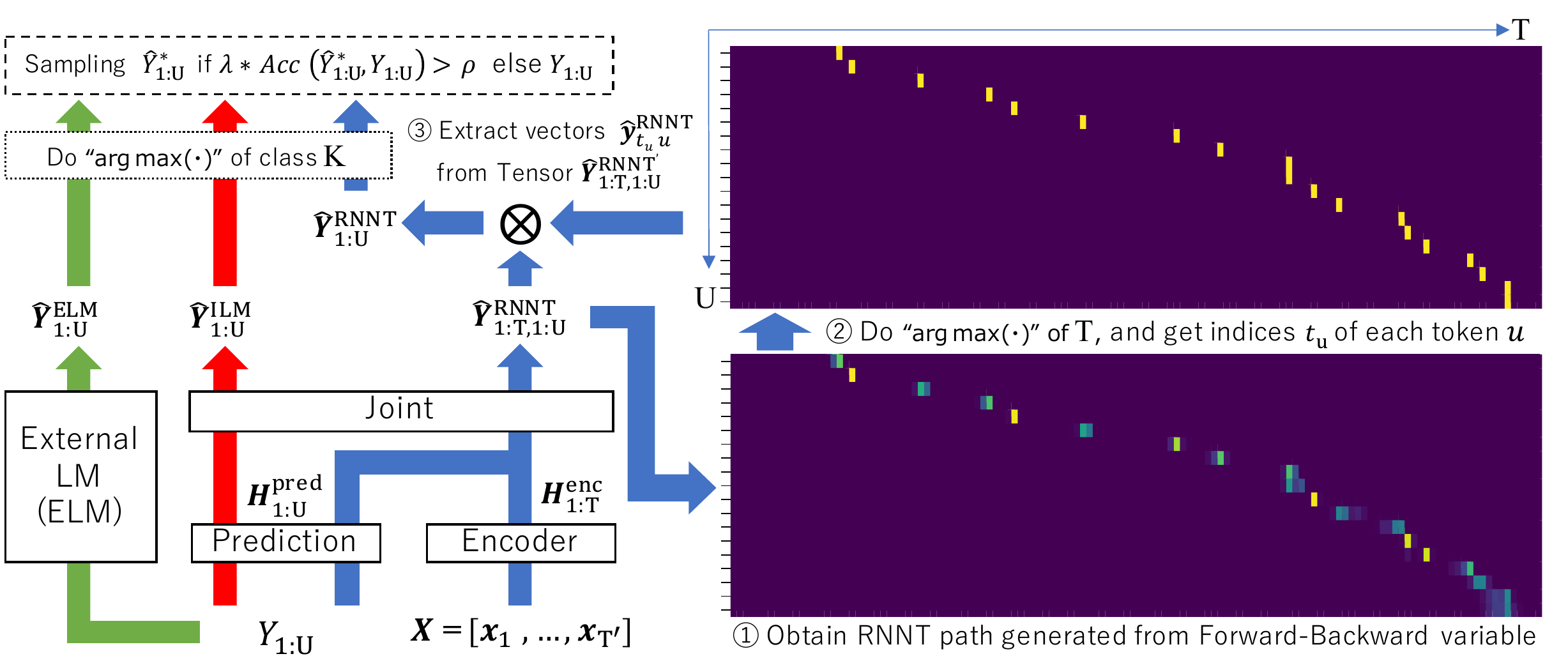}
    \vspace{-0.4cm}
    \caption{Schematic diagram of utterance-level SS. 
        The green, red and blue arrows indicate utterance-level SS procedure with ELM, ILM and RNNT, respectively. 
        The bottom and top right pictures illustrate RNNT path and time indices $t_{u}$ (yellow plots) of each token $u$, respectively. 
        Each yellow plot is the maximum arguments of time axis of RNNT path, and used to extract $\hat{\bm{y}}^{\text{RNNT}}_{t_{u}, u}$ from three dimensional tensor $\hat{\bm{Y}}^{\text{RNNT}}_{1:T, 1:U}$. 
    }
  \label{fig:uss}
  \end{center}
  \vspace{-0.8cm}
\end{figure*}

\vspace{-0.2cm}
\section{Scheduled Sampling (SS) for RNNT}
\label{sec:ss}
\vspace{-0.2cm}
SS~\cite{bengio2015ss} was originally proposed in AED training. 
SS randomly samples some tokens from actual outputs of models during training, 
and uses them to replace some ground truth tokens. 
The labels containing errors based on the training model are fed back to itself. 
Therefore, the models with SS are most robust to errors than those without SS,
and offer improved recognition performance. 

SS can be easily applied to AED, which performs LM-like training,
because the length of AED outputs and ground truth tokens is the same at length-$U$.
However, SS is difficult to apply to RNNT as its output length of-$T \times U$ differs from that of the AED output. 
Thus SS has not been fully examined for RNNT. 
In this paper, we investigate five SS approaches including four our proposals. 

\vspace{-0.2cm}
\subsection{Token-level SS using ELM/ILM}
\label{ssec:tss}
\vspace{-0.1cm}
Token-by-token SS is performed in~\cite{bengio2015ss,Cui2021rnntss},
and each token is sampled according to probability $\rho$ as follows:
\begin{equation}
  \label{eq:tss}
  \vspace{-0.1cm}
y^{\prime}_u =
    \begin{cases}
        \hat{y}^{\text{LM}}_u  &  \text{if} \ \lambda > \rho \\
        y_u        &   \text{otherwise}
    \end{cases}
    %\vspace{-0.2cm}
\end{equation}
where $\lambda$ is a hyperparameter. 
$\hat{y}^{\text{LM}}_u$ is the maximum arguments of $\hat{\bm{y}}^{\text{LM}}_u$ on class $K$ axis, 
and $\hat{\bm{y}}^{\text{LM}}_u$ is the $u$-th actual LM output.
If $\rho$ is less than $\lambda$, we replace $\hat{y}^{\text{LM}}_u$ with ground truth token $y_u$,
Note that $\rho$ is sampled from a uniform distribution. 
If LM is ELM, its token-level SS is the same as in~\cite{Cui2021rnntss}. 

The token-level SS with ELM~\cite{Cui2021rnntss} is reasonable and makes sense for RNNT, 
but its sampling does not follow the distribution of RNNT itself. 
Rather, it is similar to knowledge distillation using a pretrained ELM~\cite{bai2019aedss,kubo2022kdlm2rnnt}. 
We assume that $\hat{y}^{\text{LM}}_u$ should be sampled from a distribution conditioned on the RNNT model. 
In this paper, we propose the variant that replaces ELM with ILM so its parameters $\theta^{\text{ILM}}$ are fully contained in $\theta^{\text{RNNT}}$. 
Therefore, the sampled token $\hat{y}^{\text{LM}}_u$ follows the distribution based on RNNT. 

\vspace{-0.2cm}
\subsection{Proficieny-based utterance-level SS}
\label{ssec:uss}
\vspace{-0.1cm}
Token-level SS is performed on token-by-token, 
and thus it cannot process all tokens at once for each minibatch. 
This slows down training. 
To solve this problem, we propose utterance-level SS, see Fig.~\ref{fig:uss}. 
The utterance-level SS samples each $Y^{\prime}_{1:U}$ simultaneously for each utterance,
which raises training speed.\footnote{The total number of forwarding times of ELM and ILM for sampling process is the same as token-level SS. Note that SS using RNNT can be applied to only utterance-level variant due to the forward-backward variable computed all at once for each utterance.}
Utterance-level SS is defined as follows:
\begin{equation}  \label{eq:uss}
  \vspace{-0.2cm}
Y^{\prime}_{1:U} =
    \begin{cases}
        \hat{Y}^{\text{*}}_{1:U}   &  \text{if} \ \lambda * Acc (\hat{Y}^{\text{*}}_{1:U}, Y_{1:U}) > \rho \\
        Y_{1:U}        &   \text{otherwise}
    \end{cases}
\end{equation}
%where $\hat{Y}^{\text{*}}_{1:U} \in \mathbb{R}^{U}$ is a sequence determined by the maximum arguments of model output $\hat{\bm{Y}}^{\text{*}}_{1:U} \in \mathbb{R}^{U \times K}$ on class $K$ axis.
where $\hat{Y}^{\text{*}}_{1:U}$ is a sequence determined by the maximum arguments of model output $\hat{\bm{Y}}^{\text{*}}_{1:U} \in \mathbb{R}^{U \times K}$ on class $K$ axis.
$\hat{\bm{Y}}^{\text{*}}_{1:U}$ is obtained from ELM, ILM, or RNNT,
and its procedure is described as follows:
\begin{itemize}
  \setlength{\leftskip}{-0.6cm}
\item \noindent \textbf{Utterance-level SS using ELM/ILM:}
The green and red arrows in Fig. \ref{fig:uss} indicate utterance-level SS using ELM or ILM, respectively.
Both approaches use the same procedure as follows.
First, whole ground truth tokens are input into ELM or ILM, 
which output matrix $\hat{\bm{Y}}^{\text{ELM}}_{1:U}$ or $\hat{\bm{Y}}^{\text{ILM}}_{1:U}$, respectively.
Then, we obtain the maximum arguments of class $K$, i.e. $\hat{Y}^{\text{ELM}}_{1:U}$ or $\hat{Y}^{\text{ILM}}_{1:U}$. 

\item \noindent \textbf{Utterance-level SS using RNNT:}
The introduction of length-$T$ complicates the application of utterance-level SS to RNNT 
because the RNNT output $\hat{\bm{Y}}^{\text{RNNT}}_{1:T,1:U}$ forms a three dimensional tensor during training. 
In this paper, we propose utterance-level SS using RNNT;
the blue arrows and its description from \maru{1} to \maru{3} in Fig. \ref{fig:uss} show its procedure.
First, we obtain the RNNT paths generated from forward-backward variables, 
and then get the time index $t_{u}$ of the $u$-th non-blank token occurrence on RNNT paths.
Each $t_{u}$ is determined by the argument of the maximum on the time axis.
Finally, we extract $u$-th vector $\hat{\bm{y}}^{\text{RNNT}}_{t_u, u} \in \mathbb{R}^{K}$ from $\hat{\bm{Y}}^{\text{RNNT}}_{1:T,1:U}$, 
and gather them to make $\hat{\bm{Y}}^{\text{RNNT}}_{1:U} \in \mathbb{R}^{U \times K}$. 
The size of $\hat{\bm{Y}}^{\text{RNNT}}_{1:U}$ is the same as matrix $\hat{\bm{Y}}^{\text{LM}}_{1:U}$, 
so utterance-level SS can be applied.%\footnote{SS using RNNT can be applied to only utterance-level variant due to the forward-backward variable computed all at once for each utterance.} 
\end{itemize}

After obtaining $\hat{Y}^{\text{*}}_{1:U}$, we apply it to Eq. (\ref{eq:uss}).
However, if $\rho$ is less than constant $\lambda$ as in token-level SS, 
it is risky to replace $Y_{1:U}$ with $\hat{Y}^{\text{*}}_{1:U}$ all at once as this may result in the collapse of RNNT training. 
To mitigate this problem, we introduce the function ``$Acc ( \cdot )$'' which represents the proficiency level of each training model~\cite{moriya2020sdaed}. 
``$Acc ( \cdot )$'' treats $\hat{Y}^{\text{*}}_{1:U}$ and $Y_{1:U}$; 
it is calculated on-the-fly from the training minibatch set.
``$Acc ( \cdot )$'' ranges from 0 to 1.
At the beginning of RNNT training, ILM and RNNT may not be very accurate and ``$Acc ( \cdot )$'' is small.
This prevents training from being degraded by unreliable outputs of ILM and RNNT. 
As training progresses, these models become more accurate, 
and gradually reinforce utterance-level SS using ILM or RNNT. 

%\vspace{-0.2cm}
\section{Experiments}
\label{sec:result}
\vspace{-0.3cm}

\subsection{Data}
\label{ssec:data}
\vspace{-0.2cm}
We evaluated our proposal on three ASR tasks;
TED-LIUM release-2 (TLv2)~\cite{tedlium2}, Switchboard (SWBD)~\cite{switchboard2015}, 
and a corpus of spontaneous Japanese (CSJ)~\cite{maekawa2000}. 
The datasets contain samples totaling 210, 260 and 581 hours, respectively.
In this paper, we adopted 500 subwords, 1000 subwords, and 3262 characters 
as the output tokens for TLv2, SWBD, and CSJ, respectively.
Speed perturbation (x3)~\cite{povey2015sp} was applied to all datasets as in the ESPnet setups~\cite{espnet}. 

\vspace{-0.35cm}
\subsection{System configuration}
\label{ssec:system}
\vspace{-0.2cm}
The input feature was an 80-dimensional log Mel-filterbank appended with 3 pitch features,
and SpecAugment~\cite{specaugment} was applied.
We used a Conformer-based encoder with the same 
architecture as Conformer (L)~\cite{anmol2020conformer}; kernel size was 15.
We also replaced batch normalization with layer normalization in the convolution module. 
We used four-layer 2D-convolutional nerual networks followed by the Conformer blocks,
with the stride sizes of max-pooling layers at second and last layers of 3 and 2,
respectively.\footnote{In CSJ task, both the stride sizes in max-pooling layers were set to 2, as this was the best configuration.}
The prediction network had a 640-dimensional long short-term memory (LSTM) layer followed by a 512-dimensional feed-forward network. 
All model parameters were randomly initialized, 
and trained by using the Adam optimizer with 25k warmup for a total of 100 epochs.
$\alpha$ and $\beta$ in Eq. (\ref{eq:trainloss}) were set to 0.5 and 0.1, respectively. 

External LMs were used only for SS with ELM in the training step.
The ELMs consisted of a one-layer LSTM with 512 cells. 
The numbers of each LM output corresponded to the RNNT outputs.
Each ELM for SS was trained using the same transcribed data as RNNT. 
For decoding on TLv2 and SWBD tasks, LMs were also used; 
each consisted of a 16-block transformer-LM~\cite{moriya2020sdaed} trained with the same data as present in the ESPnet recipes. 

Hyperparameter $\lambda$ was set to 0.05, 0.15 and 0.25 for token-level SS,
and 0.15, 0.25, 0.5 for utterance-level SS.\footnote{In preliminary experiment on TLv2;
we observed ``$Acc (\hat{Y}^{\text{*}}_{1:U}, Y_{1:U})$'' using the trained ELM,
ILM and RNNT on the training set, with averaged values of 0.34, 0.41 and 0.92, respectively. 
Therefore, the actual range of $\rho$ in token- and utterance-level SS using each LM were almost the same.}
For decoding, we used alignment-length synchronous decoding~\cite{saon2020alsd} with beam size of 8.
$Z$, $\mu_{1}$, $\mu_{2}$ and $\mu_{3}$ on TLv2 and SWBD tasks were set to 1.6, 0.4, 0.2 and 0.4, respectively.
In CSJ task, we did not use any LMs, and thus $Z$ and $\mu_{3}$ were set to 1.6 and -0.4, respectively. 
We evaluated performance in terms of word error rate (WER) for English tasks,
and character error rate (CER) for Japanese tasks. 

%% TED-LIUM2 results
\begin{table}[t]
%\vspace{-0.15cm}
\centering
\caption{Comparisons of each SS for WERs on TLv2 test set.
  }
%\vspace{0.1cm}
\label{tbl:ted2}
\small
\centering
\begin{tabular}{|lccc|c|}
\hline
\multicolumn{1}{|l|}{ID}  & \multicolumn{1}{c|}{SS-level}                    & \multicolumn{1}{c|}{SS from}                & $\lambda$ & WER \\ \hline
\multicolumn{1}{|l|}{T0}  & \multicolumn{1}{c|}{-}                          & \multicolumn{1}{c|}{-}                      & -      & 7.0 \\ \hline
\multicolumn{1}{|l|}{T1}  & \multicolumn{1}{c|}{\multirow{6}{*}{token}}     & \multicolumn{1}{c|}{\multirow{3}{*}{ELM~\cite{Cui2021rnntss}}}   & 0.05   & 6.8 \\ \cline{1-1} \cline{4-5} 
\multicolumn{1}{|l|}{T2}  & \multicolumn{1}{c|}{}                           & \multicolumn{1}{c|}{}                       & 0.15   & 6.9 \\ \cline{1-1} \cline{4-5} 
\multicolumn{1}{|l|}{T3}  & \multicolumn{1}{c|}{}                           & \multicolumn{1}{c|}{}                       & 0.25   & 6.9 \\ \cline{1-1} \cline{3-5} 
\multicolumn{1}{|l|}{T4}  & \multicolumn{1}{c|}{}                           & \multicolumn{1}{c|}{\multirow{3}{*}{ILM}}   & 0.05   & 6.8 \\ \cline{1-1} \cline{4-5} 
\multicolumn{1}{|l|}{T5}  & \multicolumn{1}{c|}{}                           & \multicolumn{1}{c|}{}                       & 0.15   & 6.6 \\ \cline{1-1} \cline{4-5} 
\multicolumn{1}{|l|}{T6}  & \multicolumn{1}{c|}{}                           & \multicolumn{1}{c|}{}                       & 0.25   & 6.8 \\ \hline
\multicolumn{1}{|l|}{T7}  & \multicolumn{1}{c|}{\multirow{9}{*}{utterance}} & \multicolumn{1}{c|}{\multirow{3}{*}{ELM}}   & 0.15   & 6.8 \\ \cline{1-1} \cline{4-5} 
\multicolumn{1}{|l|}{T8}  & \multicolumn{1}{c|}{}                           & \multicolumn{1}{c|}{}                       & 0.25   & 6.8 \\ \cline{1-1} \cline{4-5} 
\multicolumn{1}{|l|}{T9}  & \multicolumn{1}{c|}{}                           & \multicolumn{1}{c|}{}                       & 0.50   & 6.6 \\ \cline{1-1} \cline{3-5} 
\multicolumn{1}{|l|}{T10} & \multicolumn{1}{c|}{}                           & \multicolumn{1}{c|}{\multirow{3}{*}{ILM}}   & 0.15   & 6.8 \\ \cline{1-1} \cline{4-5} 
\multicolumn{1}{|l|}{T11} & \multicolumn{1}{c|}{}                           & \multicolumn{1}{c|}{}                       & 0.25   & 6.7 \\ \cline{1-1} \cline{4-5} 
\multicolumn{1}{|l|}{\textbf{T12}} & \multicolumn{1}{c|}{}                           & \multicolumn{1}{c|}{}                       & 0.50   & \textbf{6.5} \\ \cline{1-1} \cline{3-5} 
\multicolumn{1}{|l|}{T13} & \multicolumn{1}{c|}{}                           & \multicolumn{1}{c|}{\multirow{3}{*}{RNNT}} & 0.15   & 6.6 \\ \cline{1-1} \cline{4-5} 
\multicolumn{1}{|l|}{T14} & \multicolumn{1}{c|}{}                           & \multicolumn{1}{c|}{}                       & 0.25   & 6.7 \\ \cline{1-1} \cline{4-5} 
\multicolumn{1}{|l|}{T15} & \multicolumn{1}{c|}{}                           & \multicolumn{1}{c|}{}                       & 0.50   & 6.7 \\ \hline
%\multicolumn{4}{|l|}{}                                                                                                             &     \\
\multicolumn{4}{|l|}{Our best system (T12) + Transformer-LM}                                                                                                             & 6.0    \\ \hline \hline
\multicolumn{4}{|l|}{Hybrid NN-HMM + 4-gram LM~\cite{zhou2020hmmsota}}                                                                                                             & 7.3    \\ \cline{5-5}
\multicolumn{4}{|l|}{+ Transformer-LM~\cite{zhou2020hmmsota}}                                                                                              	       & 5.6    \\ \hline
\multicolumn{4}{|l|}{Phoneme RNNT + 4-gram LM~\cite{zhou2021rnntcess}}                                                                                                             & 6.9   \\ \cline{5-5}
\multicolumn{4}{|l|}{+ LSTM-LM + Transformer-LM~\cite{zhou2021rnntcess}}                                                                                                             & 6.0    \\ \hline
\end{tabular}
\vspace{-0.35cm}
\end{table}

%% Switchboard results
\begin{table}[t]
%\vspace{-0.15cm}
\centering
\caption{Comparisons of each SS for WERs on SWBD hub5'00 set.
  }
%\vspace{0.1cm}
\label{tbl:swbd}
\small
\centering
\hspace{-0.3cm}
\begin{tabular}{|lccc|ccc|}
\hline
\multicolumn{1}{|l|}{\multirow{2}{*}{ID}} & \multicolumn{1}{c|}{\multirow{2}{*}{SS-level}}   & \multicolumn{1}{c|}{\multirow{2}{*}{SS from}} & \multirow{2}{*}{$\lambda$} & \multicolumn{3}{c|}{WER}                                     \\ \cline{5-7} 
\multicolumn{1}{|l|}{}                    & \multicolumn{1}{c|}{}                           & \multicolumn{1}{c|}{}                         &                         & \multicolumn{1}{c|}{SWB} & \multicolumn{1}{c|}{CH}   & $\sum$  \\ \hline
\multicolumn{1}{|l|}{S0}                  & \multicolumn{1}{c|}{-}                          & \multicolumn{1}{c|}{-}                        & -                       & \multicolumn{1}{c|}{6.9}  & \multicolumn{1}{c|}{14.4} & 10.6 \\ \hline
\multicolumn{1}{|l|}{S1}                  & \multicolumn{1}{c|}{\multirow{6}{*}{token}}     & \multicolumn{1}{c|}{\multirow{3}{*}{ELM~\cite{Cui2021rnntss}}}     & 0.05                    & \multicolumn{1}{c|}{6.7}  & \multicolumn{1}{c|}{13.8} & 10.3 \\ \cline{1-1} \cline{4-7} 
\multicolumn{1}{|l|}{S2}                  & \multicolumn{1}{c|}{}                           & \multicolumn{1}{c|}{}                         & 0.15                    & \multicolumn{1}{c|}{6.8}  & \multicolumn{1}{c|}{13.8} & 10.3 \\ \cline{1-1} \cline{4-7} 
\multicolumn{1}{|l|}{S3}                  & \multicolumn{1}{c|}{}                           & \multicolumn{1}{c|}{}                         & 0.25                    & \multicolumn{1}{c|}{6.7}  & \multicolumn{1}{c|}{13.8} & 10.3 \\ \cline{1-1} \cline{3-7} 
\multicolumn{1}{|l|}{S4}                  & \multicolumn{1}{c|}{}                           & \multicolumn{1}{c|}{\multirow{3}{*}{ILM}}     & 0.05                    & \multicolumn{1}{c|}{6.7}  & \multicolumn{1}{c|}{13.6} & 10.2 \\ \cline{1-1} \cline{4-7} 
\multicolumn{1}{|l|}{S5}                  & \multicolumn{1}{c|}{}                           & \multicolumn{1}{c|}{}                         & 0.15                    & \multicolumn{1}{c|}{6.7}  & \multicolumn{1}{c|}{13.7} & 10.2 \\ \cline{1-1} \cline{4-7} 
\multicolumn{1}{|l|}{S6}                  & \multicolumn{1}{c|}{}                           & \multicolumn{1}{c|}{}                         & 0.25                    & \multicolumn{1}{c|}{6.6}  & \multicolumn{1}{c|}{13.7} & 10.2 \\ \hline
\multicolumn{1}{|l|}{S7}                  & \multicolumn{1}{c|}{\multirow{9}{*}{utterance}} & \multicolumn{1}{c|}{\multirow{3}{*}{ELM}}     & 0.15                    & \multicolumn{1}{c|}{6.7}  & \multicolumn{1}{c|}{14.1} & 10.3 \\ \cline{1-1} \cline{4-7} 
\multicolumn{1}{|l|}{S8}                  & \multicolumn{1}{c|}{}                           & \multicolumn{1}{c|}{}                         & 0.25                    & \multicolumn{1}{c|}{6.7}  & \multicolumn{1}{c|}{14.2} & 10.4 \\ \cline{1-1} \cline{4-7} 
\multicolumn{1}{|l|}{S9}                  & \multicolumn{1}{c|}{}                           & \multicolumn{1}{c|}{}                         & 0.50                    & \multicolumn{1}{c|}{\textbf{6.5}}  & \multicolumn{1}{c|}{13.9} & 10.2 \\ \cline{1-1} \cline{3-7} 
\multicolumn{1}{|l|}{\textbf{S10}}                 & \multicolumn{1}{c|}{}                           & \multicolumn{1}{c|}{\multirow{3}{*}{ILM}}     & 0.15                    & \multicolumn{1}{c|}{6.7}  & \multicolumn{1}{c|}{\textbf{13.3}} & \textbf{10.0} \\ \cline{1-1} \cline{4-7} 
\multicolumn{1}{|l|}{S11}                 & \multicolumn{1}{c|}{}                           & \multicolumn{1}{c|}{}                         & 0.25                    & \multicolumn{1}{c|}{6.8}  & \multicolumn{1}{c|}{13.6} & 10.2 \\ \cline{1-1} \cline{4-7} 
\multicolumn{1}{|l|}{S12}                 & \multicolumn{1}{c|}{}                           & \multicolumn{1}{c|}{}                         & 0.50                    & \multicolumn{1}{c|}{6.8}  & \multicolumn{1}{c|}{13.9} & 10.4 \\ \cline{1-1} \cline{3-7} 
\multicolumn{1}{|l|}{S13}                 & \multicolumn{1}{c|}{}                           & \multicolumn{1}{c|}{\multirow{3}{*}{RNNT}}   & 0.15                    & \multicolumn{1}{c|}{6.7}  & \multicolumn{1}{c|}{13.8} & 10.3 \\ \cline{1-1} \cline{4-7} 
\multicolumn{1}{|l|}{S14}                 & \multicolumn{1}{c|}{}                           & \multicolumn{1}{c|}{}                         & 0.25                    & \multicolumn{1}{c|}{6.6}  & \multicolumn{1}{c|}{13.6} & 10.1 \\ \cline{1-1} \cline{4-7} 
\multicolumn{1}{|l|}{S15}                 & \multicolumn{1}{c|}{}                           & \multicolumn{1}{c|}{}                         & 0.50                    & \multicolumn{1}{c|}{6.6}  & \multicolumn{1}{c|}{13.6} & 10.1 \\ \hline
\multicolumn{4}{|l|}{Our best system (S10) + Transformer-LM}                                                                                                                                                & \multicolumn{1}{c|}{6.2}     & \multicolumn{1}{c|}{12.3}     & 9.3     \\ \hline \hline
\multicolumn{4}{|l|}{Conformer-AED~\cite{tsunke2021caedsota}}                                                                                                                                                & \multicolumn{1}{c|}{6.7}     & \multicolumn{1}{c|}{13.0}     & 9.9     \\ \cline{5-7} 
\multicolumn{4}{|l|}{+ LSTM-LM + Transformer-XL LM~\cite{tsunke2021caedsota}}                                                                                                                                                & \multicolumn{1}{c|}{5.5}     & \multicolumn{1}{c|}{11.2}     & 8.4     \\ \hline %\cline{5-7}
%\multicolumn{4}{|l|}{+ System combination~\cite{tsunke2021caedsota}}                                                                                                                                                & \multicolumn{1}{c|}{5.0}     & \multicolumn{1}{c|}{10.0}     & -     \\ \hline
\multicolumn{4}{|l|}{Phoneme RNNT + 4-gram LM~\cite{zhou2022phonemernnt}}                                                                                                                                                & \multicolumn{1}{c|}{-}     & \multicolumn{1}{c|}{-}     & 10.3     \\ \cline{5-7} 
\multicolumn{4}{|l|}{+ Transformer-LM~\cite{zhou2022phonemernnt}}                                                                                                                                                & \multicolumn{1}{c|}{-}     & \multicolumn{1}{c|}{-}     & 9.2 \\ \hline
\multicolumn{4}{|l|}{RNNT~\cite{Cui2021rnntss}}                                                                                                                                                & \multicolumn{1}{c|}{6.9}     & \multicolumn{1}{c|}{14.9}     & 10.9     \\ \cline{5-7} 
\multicolumn{4}{|l|}{RNNT + LSTM-LM~\cite{Cui2021rnntss}}                                                                                                                                                & \multicolumn{1}{c|}{6.0}     & \multicolumn{1}{c|}{13.0}     & 9.5     \\ \hline
\end{tabular}
\vspace{-0.25cm}
\end{table}

%% CSJ results
\begin{table}[t]
%\vspace{-0.15cm}
\centering
\caption{Comparisons of each SS for WERs on CSJ evaluation sets.
  }
%\vspace{0.1cm}
\label{tbl:csj}
\small
\centering
\hspace{-0.2cm}
\begin{tabular}{|lccc|c|ccc|}
\hline
\multicolumn{1}{|l|}{\multirow{2}{*}{ID}} & \multicolumn{1}{c|}{\multirow{2}{*}{SS-level}}   & \multicolumn{1}{c|}{\multirow{2}{*}{SS from}} & \multirow{2}{*}{$\lambda$} & \multirow{2}{*}{\#Param} & \multicolumn{3}{c|}{CER}                                  \\ \cline{6-8} 
\multicolumn{1}{|l|}{}                    & \multicolumn{1}{c|}{}                           & \multicolumn{1}{c|}{}                         &                         &                          & \multicolumn{1}{c|}{E1}  & \multicolumn{1}{c|}{E2}  & E3  \\ \hline
\multicolumn{1}{|l|}{C0}                  & \multicolumn{1}{c|}{-}                          & \multicolumn{1}{c|}{-}                        & -                       & \multirow{4}{*}{119M}    & \multicolumn{1}{c|}{4.2} & \multicolumn{1}{c|}{3.2} & 3.6 \\ \cline{1-4} \cline{6-8} 
\multicolumn{1}{|l|}{C1}                  & \multicolumn{1}{c|}{\multirow{3}{*}{utterance}} & \multicolumn{1}{c|}{ELM}                      & 0.25                    &                          & \multicolumn{1}{c|}{4.0} & \multicolumn{1}{c|}{\textbf{2.9}} & 3.4 \\ \cline{1-1} \cline{3-4} \cline{6-8} 
\multicolumn{1}{|l|}{\textbf{C2}}                  & \multicolumn{1}{c|}{}                           & \multicolumn{1}{c|}{ILM}                      & 0.25                    &                          & \multicolumn{1}{c|}{\textbf{3.9}} & \multicolumn{1}{c|}{3.0} & \textbf{3.2} \\ \cline{1-1} \cline{3-4} \cline{6-8} 
\multicolumn{1}{|l|}{C3}                  & \multicolumn{1}{c|}{}                           & \multicolumn{1}{c|}{RNNT}                    & 0.25                    &                          & \multicolumn{1}{c|}{\textbf{3.9}} & \multicolumn{1}{c|}{3.0} & 3.3 \\ \hline
\multicolumn{4}{|l|}{System Combination (C1+C2+C3)}                                                                                                                                                & -                     & \multicolumn{1}{c|}{3.8} & \multicolumn{1}{c|}{2.8} & 3.1 \\ \hline \hline
\multicolumn{4}{|l|}{Conformer-Transducer~\cite{karita2021csj}}                                                                                                                                                & 120M                     & \multicolumn{1}{c|}{4.1} & \multicolumn{1}{c|}{3.2} & 3.5 \\ \hline
\end{tabular}
\vspace{-0.35cm}
\end{table}

\vspace{-0.3cm}
\subsection{Results}
\label{ssec:results}
\vspace{-0.2cm}

First, we investigate the effectiveness of each SS in improving RNNT performance on TLv2 task; 
the conditions and resulting WERs are shown in Table~\ref{tbl:ted2}.
``SS-level'' indicates sampling method, i.e. token-level SS (``token'') or utterance-level SS (``utterance'').
``SS-from'' means the sampling model which is ELM, ILM or RNNT,
and $\lambda$ is threshold parameter in Eq. (\ref{eq:tss}) or (\ref{eq:uss}). 
``T0'' without SS is baseline, and the other systems used SS. 
LM was not used during inferencing for all results ``T*''. 
The bottom block of Table~\ref{tbl:ted2} shows the results from the literature.
We can see that the systems trained with SS yielded better WERs than the baseline.
In particular, the models trained with utterance-level SS achieved better WERs than those with token-level SS. 
System ``T12'' which uses utterance-level SS with ILM and $\lambda = 0.5$ achieved the best performance. 
``T12'' was further improved by using Transformer-LM. 
Those WERs are competitive with the transducer-based state-of-the-art although our system did not use n-gram LM. 

We also evaluated our proposed SS on the SWBD task; 
the results are shown in Table~\ref{tbl:swbd}.
``S0'' without SS is baseline, and the other systems used SS.
We can see that the model trained with SS achieved better WERs than the baseline.
System ``S10'' which uses utterance-level SS with ILM and $\lambda = 0.15$ achieved the best performance,
and was only slightly behind state-of-the-art E2E ASR system~\cite{tsunke2021caedsota} in the bottom block of Table~\ref{tbl:swbd}. 
``S10'' was further improved by using Transformer-LM, 
but still there was some gap between our best system and state-of-the-art WERs.\footnote{We also evaluated our best RNNT (E10) w/o and w/ Transformer-LM on hub5'01/Rt03. The WERs of swb/swb2p3/swb2p4/$\sum$ on hub5'01 were 6.9/9.7/13.6/10.2 (w/ Transformer-LM: 6.3/8.9/12.4/9.3). The WERs of swb/fsh/$\sum$ on Rt03 were 14.9/9.5/12.3 (w/ Transformer-LM: 13.3/8.5/11.0).} 

Finally, we also evaluated our proposals on CSJ dataset. 
The results are shown in Table~\ref{tbl:csj}.
``C0'' without SS is baseline, and the other systems used SS.
We only applied utterance-level SS in model training because it always outperformed token-level SS in the above experiments.
We can see that system ``C2'' with ILM achieved the best CERs. 
In particular, ``C2'' outperformed the previous state-of-the-art~\cite{karita2021csj} while keeping the model size reasonable. 
%We can see that system ``C2'' with ILM achieved the best CERs, 
%and it outperformed the previous state-of-the-art~\cite{karita2021csj} while keeping the model size reasonable.
Moreover, system combination of ``C1'', ``C2'' and ``C3'' further improved the CERs. 

%\vspace{-0.325cm}
\vspace{-0.2cm}
\section{Conclusion}
\label{ssec:conclusions}
%\vspace{-0.275cm}
\vspace{-0.2cm}
We have proposed SS variants suitable for RNNT, they do not need any external alignments or pretrained ELMs. 
The tokens are randomly sampled from ILM or RNNT outputs. 
The sampled tokens are replaced with some or all ground truth tokens for token-level or utterance-level SS, respectively. 
Then the token sequence containing errors is fed back to RNNT during training. 
The proposed variants were evaluated on three datasets, 
and found to consistently yield superior ASR performance to strong RNNT baselines without SS. 
In particular, our proficiency-based utterance-level SS with ILM achieved the best performance on all tasks. 
Moreover, our best system on CSJ outperformed the previous state-of-the-art. 
Future work includes improving utterance-level SS using RNNT with high proficiency to increase sampling variations as in~\cite{Cui2021rnntss}. 

\clearpage

\bibliographystyle{IEEEbib}
\small
\bibliography{strings,refs}
\end{document}